\documentstyle[12pt,epsf]{article}
\newlength{\overeqskip}
\newlength{\undereqskip}
\newcommand{\nc}{\newcommand}

\nc{\nn}{\nonumber \\*}

\nc{\be}{\begin{equation}}
\nc{\ee}{\end{equation}}
\nc{\ba}{\begin{eqnarray}}
\nc{\ea}{\end{eqnarray}}

\nc{\ra}{\rightarrow}
\nc{\lra}{\leftrightarrow}

\nc{\eg}{{\em e.g.} }
\nc{\etal}{{\em et.al.}}
\nc{\sgL}{{\it sign}($L$)}

\nc{\half}{{\frac{1}{2}}}
\nc{\dt}{\frac{d}{dt}}
\nc{\neut}[1]{{\nu_{#1}}}
\nc{\aneut}[1]{{\bar{\nu}_{#1}}}

\nc{\herm}{^{\dagger}}
\nc{\pri} {^{\prime}}
\nc{\ev} {\; \mbox{eV}}
\nc{\mev}{\; \mbox{MeV}}
\nc{\pp} [1]{{P_{#1}^+}}
\nc{\pmm}[1]{{P_{#1}^-}}

\def\lsim{\;\raise0.3ex\hbox{$<$\kern-0.75em \raise-1.1ex\hbox{$\sim$}}\;}
\def\gsim{\;\raise0.3ex\hbox{$>$\kern-0.75em \raise-1.1ex\hbox{$\sim$}}\;}

\nc{\bi}{{\bibitem}}
\def\eV{{\rm\ eV}}

\def\MeV{{\rm\ MeV}}

\begin{document}
%
%
\begin{titlepage}
\pagestyle{empty}
\baselineskip=21pt
\rightline{HIP-1999-39/TH}
\rightline{NORDITA-99/40 HE}
\rightline{June 21, 1999}
\vskip .6in

\begin{center} {\Large{\bf On chaoticity of the amplification of the \\
                           neutrino asymmetry in the early universe}}

\end{center}
\vskip .3in

\begin{center}

Kari Enqvist$^{1,2}$,  Kimmo Kainulainen$^3$ and Antti Sorri$^1$\\

\vskip .2in

$^1${\it  Physics Department, University of Helsinki, \\
          P.O.\ Box 9, FIN-00014 University of Helsinki }\\
\vskip .1in
$^2${\it  Helsinki Institute of Physics\\
          P.O.\ Box 9, FIN-00014 University of Helsinki }\\
\vskip .1in
$^3${\it  NORDITA, Blegdamsvej 17, DK-2100 Copenhagen \O, Denmark}\\

\end{center}

\vskip 0.3in

\centerline{ {\bf Abstract} }
\baselineskip=18pt
\vskip 0.5truecm\noindent
We consider numerically the
growth of neutrino asymmetry in active-sterile neutrino oscillations
in the early universe. It is shown that the final sign of the asymmetry can
be highly sensitive to small variations of the oscillation parameters. We
find regions which are completely or partially chaotic, but also regions
where the sign remains very robust. The consequences for atmospheric neutrino
oscillations and primordial nucleosynthesis are then discussed. In the
completely chaotic region the predicted $^4$He-abundance has an inherent
arbitrariness $\Delta Y \simeq 10^{-2}$.\\

\end{titlepage}

\baselineskip=20pt


Active-sterile neutrino oscillations in the early Universe is a fascinating
possibility with far-reaching consequences \eg\ for nucleosynthesis
[1-12]
and CMB radiation \cite{CMB}. Nucleosynthesis considerations in
particular have made it possible to place stringent constraints on
model building aimed at understanding  the observed neutrino
anomalies in terrestrial observations.  In the very first papers
\cite{dkdb} it was observed that the mixing with an active
species ($SU(2)$-doublet) endows the sterile ($SU(2)$-singlet)
neutrino with effective interactions, which can be strong enough
to bring the sterile species in equilibrium. The ensuing excess
energy density would result in a failure of the  nucleosynthesis
explanation of the observed light element abundances \cite{RecNS}.
This line of reasoning was put to a solid computational foundation
in refs.\ \cite{ekmL,ektBig}, and these results were later
reproduced in ref.\  \cite{ssf}. Of particular interest was
the observation that nucleosynthesis is in conflict with
$\nu_\mu -\nu_s$-oscillation solution to the atmospheric
neutrino problem \cite{ektat}.

Already in \cite{ekmL} it was noted that nucleosynthesis constraints
\cite{ektBig,ssf} depend on the reasonable assumption that the leptonic
asymmetries are not many orders of magnitudes larger than the baryonic
one.  More specifically, \eg\ for a mass squared difference $|\delta m^2|
= 10^{-4} \ev^2$ a large initial asymmetry, $L_{\nu}^{\rm in} \gsim 10^{-5}$
(here $L_{\alpha}=(N_{\alpha}-N_{\bar{\alpha}})/N_\gamma$) would suppress
the effective mixing angle so much that the equilibration would never
take place.  This observation was later revived by Foot and Volkas
\cite{fvL}, who basing on this and another previously observed
effect, an exponential growth of leptonic asymmetry \cite{dk},
suggested an interesting way to circumvent the nucleosynthesis
constraints without invoking unnatural initial conditions \cite{fvA}.
Their scenario assumes a novel mass-mixing scenario, where a
$\nu_\tau -\nu_{s}$-mixing, with carefully chosen parameters,
produces a large leptonic asymmetry (but does not equilibrate
$\nu_s$), which suppresses the subsequent $\nu_\mu-\nu_s$-mixing
angle and thereby prevents the $\nu_s$-equilibration from taking
place.  Some details concerning the growth of the asymmetry in
this scenario are still under debate \cite{shifu,fvD}.

It was later observed by Shi \cite{shi} that the period of exponential
growth exhibits chaotic features and therefore, while the amplitude of
the final asymmetry is robust, its {\em sign} appeared to be essentially
arbitrary. This raises some interesting questions: for example, is the
sign of $L_{\nu}$ sensitive to the fluctuations in the initial conditions,
like in the baryon asymmetry?  If so, one should expect a large
suppression in the effective asymmetry present at the important
epoch for the $\nu_\mu-\nu_s$-oscillations due to diffusion effects.
For this purpose it is important to establish the extent of possible
chaotic or regular regions in the parameter space.  In this letter we
have studied the dependence of the sign of $L_{\nu}$ on  neutrino mixing
parameters.  We find a rather clearcut division of the parameter space
into non chaotic and partly or completely chaotic\footnote{We do not
claim here that the system exhibits chaoticity
in the mathematical sense of the definition of chaos; we merely mean
that the system is sensitive to small variations of the parameters.}
regions. In chaotic regions the final sign of the asymmetry is indeed found
to be highly sensitive also to fluctuations in the initial conditions.

Another, more direct consequence follows from the fact that \sgL\ affects
the computed $^4$He abundance, either directly in the case of $\nu_e-\nu_s$
oscillations, or when induced by large $L_{\nu}$ created in $\nu_\mu-\nu_s$
or $\nu_\tau-\nu_s$ oscillations and later transferred to $\nu_e$-sector via
active-active oscillations \cite{eAsymEf} . It then follows that possible
chaotic behavior will constrain our chances to draw any definite conclusions
about the effects of sterile neutrinos on Big Bang nucleosynthesis, as
we will discuss below.\\

In the early Universe neutrinos experience frequent scatterings,
which tend to bring their distributions into thermal equilibrium.
The requisite mathematical formalism is therefore very different
from the one particle approach valid for description of accelerator
physics (beams) and even solar neutrinos. Indeed, the objects of
interest are the (reduced) density matrices for the neutrino and
antineutrino ensembles
\be
\rho_{\nu}      \equiv \half      P_0 (1 + {\bf      P}) \; , \qquad
\rho_{\bar \nu} \equiv \half \bar P_0 (1 + {\bf \bar P}).
\label{rho}
\ee
Solving full momentum dependent kinetic equations for $\rho_\nu (p)$ and
$\rho_{\bar \nu}(p)$ \cite{ektBig,pdep,stodo,georg2} is obviously a very
difficult task. Instead, we employ the momentum averaged equations for
${\bf P} = {\bf P}(\langle p \rangle )$, with
$\langle p\rangle \simeq 3.15 T$, which should be expected to give a
good approximation for the full system \cite{ektBig}. (Our preliminary
studies with full momentum dependent equations support this assumption).
Moreover, for the parameters we are interested in, one can neglect the
collision terms so that $P_0$ remains a constant and can be set to a
unity. The coupled equations of motion then are (for definiteness we
shall focus here on $\nu_\tau - \nu_{s}$ oscillations; other cases are
obtained from this by simple redefinitions\footnote{Interested
reader can find these redefinitions for example
from \cite{ektBig}.})
\ba
\dot{\bf P}
    &=& {\bf      V}\times{\bf      P}- D {\bf      P}_T \nn
\dot{\bf {\bar P}}
    &=& {\bf \bar V}\times{\bf \bar P}- D {\bf \bar P}_T
\label{one state}
\ea
where $\dot{\bf P} \equiv {d{\bf P}/dt}$ and we defined
${\bf P}_T \equiv P_x {\bf \hat x} +  P_y {\bf \hat y}$. In the case
of $\nu_\tau -\nu_{s}$ oscillations the damping coefficient is
$  D \simeq 1.8 G_F^2 T^5$ \cite{ektBig}, and $\bar D \simeq D$
to a very high accuracy. It is convenient to decompose the rotation
vector ${\bf V}$ as
\be
  {\bf V} = V_x \;{\bf \hat{x}} + \bigl( V_0 + V_L \bigr) \; {\bf \hat{z}},
\ee
with the components
\ba
  V_x  &=&  \Delta \sin 2 \theta             \nn
  V_0  &=& -\Delta \cos 2 \theta + \delta V_{\tau}  \nn
  V_L  &=&  \sqrt{2} G_F N_{\gamma} \; L,
\label{Vcomp}
\ea
where $\theta$ is the vacuum mixing angle, $\Delta \equiv \delta m^2/2
\langle p\rangle $ and the photon number density $N_\gamma \equiv 2\zeta (3)
T^3/\pi^2$. The effective asymmetry $L$ appearing in the leading contribution
$V_L$ to the neutrino effective potential is
\be
   L = - \half L_n + L_{\neut{e}} + L_{\neut{\mu}}
                    + 2 L_{\neut{\tau}}(P)
\label{alla}
\ee
where $n$ refers to neutrons and we have assumed electrical neutrality of
the plasma. The remaining piece to the effective potential $\delta V_{\tau}$
is given by \cite{ektBig,gio}
\be
  \delta V_{\tau}
    = - \sqrt{2} G_F N_\gamma A_\tau \frac{\langle p \rangle T}{2 M_Z^2},
\label{effpot}
\ee
with $A_\tau = 14\zeta(4)/\zeta(3) \simeq 12.61$.  The rotation vector
for antineutrinos is simply ${\bf \bar V}(L) = {\bf V}(-L)$. The
coupling of particle and antiparticle sectors occurs through the asymmetry
term, where
\be
L_{\nu_\tau}(P) = L_{\nu_\tau}^{\rm in} + \frac{3}{8} (P_z - \bar P_z).
\label{LofP}
\ee
with $L_{\nu_\tau}^{\rm in}$ being the initial $\neut{\tau}$ asymmetry.

Even the simplified one state-quantum kinetic equations (\ref{one state})
are very difficult to handle numerically, because of the vast difference in
the time scales involved (Hubble expansion rate, matter
oscillation frequency and the width of the resonance, for example) on one hand
and due to extremely strong coupling induced by
the asymmetry term on the other.
The so-called static approximation employed in \cite{fvL} reduces the system
to one first order differential equation. Unfortunately it is not really
suitable for the treatment of oscillations at the resonance, since many of
its basic assumptions -- that the system is adiabatic, that the MSW-effect
can be neglected, and that the rate of change of lepton number is dominated
by the collisions -- break down at the resonance.

These considerations emphasize the need for a very careful numerical
approach. In practice, the accuracy is much improved if one makes separation
between the large ($\sim$ number density) and small components ($\sim$
asymmetry) in equations (\ref{one state}). To this end we change
the variables into
\be
 P_i^\pm \equiv P_i \pm \bar{P}_i,
\ee
in terms of which (\ref{one state}) become
\def\phm{\phantom{-}}
\begin{eqnarray}
\dot{P}_x^+ & = &   -  V_0 P_y^+  - V_L P_y^- - D P_x^+ \nn
\dot{P}_y^+ & = & \phm V_0 P_x^+  + V_L P_x^- - D P_y^+ \nn
\dot{P}_z^+ & = & \phm V_x P_y^+                        \nn
\dot{P}_x^- & = &   -  V_0 P_y^-  - V_L P_y^+ - D P_x^- \nn
\dot{P}_y^- & = & \phm V_0 P_x^-  + V_L P_x^+ - D P_y^- \nn
\dot{P}_z^- & = & \phm V_x P_y^- .
\label{neweqs}
\end{eqnarray}
We have studied numerically the behaviour of the system described by
(\ref{neweqs}) as a function of the oscillation parameters $\delta m^2$
and $\sin^22\theta$, and in particular the evolution of the asymmetry
(\ref{LofP}). Of crucial importance in this evolution is the occurrence
of the resonance at $V_0 = 0$ if  $\delta m^2 < 0$. Inserting the
appropriate parameters, one finds that the resonance temperature is
given by \cite{ektBig}
\be
T_{\rm res} \simeq 16.0 \; (|\delta m^2| \cos 2\theta)^{1/6} \; \MeV,
\label{Tres}
\ee
where $\delta m^2$ is given in units $\eV^2$. Far above the resonance
the damping terms tend to suppress the off-diagonal elements ${\bf P}_T$
and moreover, the system is driven towards the initially stable fixed
point $L = 0$.  As soon as the system passes the resonance however, $L=0$
becomes an unstable fixed point and two new locally stable and degenerate
minima corresponding to the solutions of $V_L + V_0 = 0$ appear;
these are given by the condition
    $|L| \simeq 11 (|\delta m^2|/{\rm eV}^2)\,({\rm MeV}/T)^4$.

The system is roughly analogous to a ball rolling down a valley that branches
to two, passing via a saddle configuration. Initially the  branching into
these
two new valleys can be very shallow and it may stay that for a long time. Once
on the side of the bifurcation, the ball still keeps on passing over the
central
barrier (the continuation of the old stable fixed point to an  unstable
extremum)
until the barrier grows too high, or until friction (damping) reduces energy
enough, and it gets trapped to one of the valleys.  It is easy to picture in
ones mind that in a case of a very shallow bifurcation a small change in the
initial conditions ($L^{\rm in}$), or in the shape of the valleys (oscillation
parameters), can very much affect which minimum the system finally chooses to
settle in.
\\
\begin{figure}[t]
\centering
\vspace*{-10mm}
\hspace*{-0mm}
\leavevmode\epsfysize=13.0cm \epsfbox{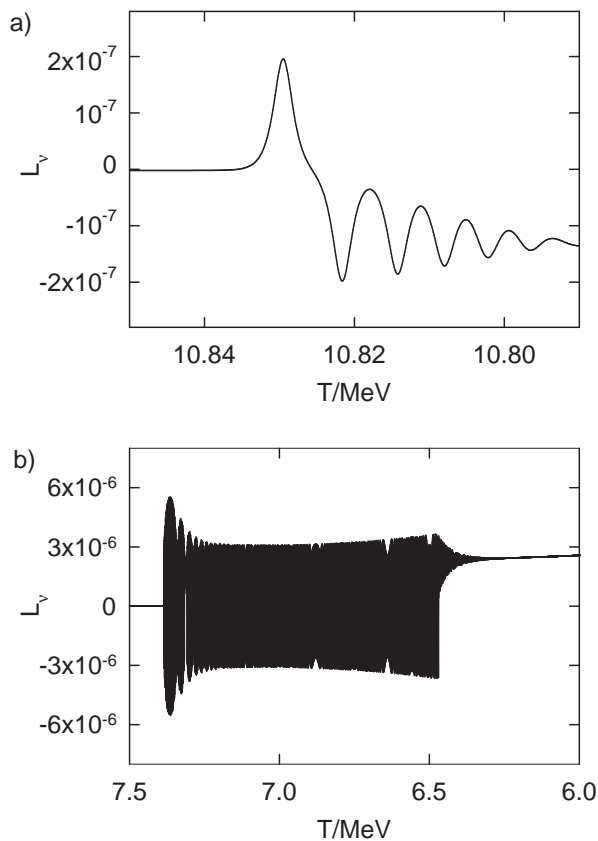}\\[-10mm]
\caption[fig2]{
         The evolution of the $\nu_\tau$-asymmetry in the neigbourhood
         of the resonance. In Fig. 1a
         $\delta m^2 =   -10^{-1}$ and
         $\sin^22\theta = 2 \times 10^{-9}$; in
         Fig. 1b
         $\delta m^2 =  -10^{-2}$ and
         $\sin^22\theta = 2 \times 10^{-5}$. }
\end{figure}

In Fig.\ 1 we show the asymmetry $L_{\nu_\tau}$ (\ref{LofP}) resulting
from solving equations (\ref{neweqs}) for two sets of parameters. In Fig.\
1a the resonance is rather narrow and only one
oscillation occurs before the system is trapped into the minimum with
a negative sign of $L_{\nu_\tau}$.  The subsequent oscillations about
this new local minimum are quickly washed away by the damping terms.
In contrast, Fig.\ 1b shows an example of oscillation parameters with which
the bifurcation into new local minimum is extremely slow and for a long
time there is hardly any barrier between the two minima with opposite
signs of $L_{\nu_\tau}$, and the system oscillates thousands
of times before settling down to a minimum with positive
$L_{\nu_\tau}$. After settling down, the further evolution of the
asymmetry follows a power-like behaviour. These results agree well
with those of ref.\ \cite{shi}.

It is instructive to look a little more carefully into how the system
approaches the resonance. Before the resonance the off-diagonal $P^\pm$
components are very near zero and $\pmm{z}$ near the value $L=0$. Just
before the resonance the $P^+_{x,y}$ components begin to increase, which
triggers both the growth of $P^-_{x,y}$ components and the decrease of
$P_z^+$. As $V_0$ changes the sign at the resonance it creates an
instability in the equation for $P_y^-$, which eventually strongly pushes
$P_y^-$ to negative direction. The simple coupling of $P_y^-$ to $P_z^-$
in (\ref{neweqs}) then drags $P_z^-$ along leading to a rapid growth of
$L_{\nu_\tau}$.  So far these phenomena have not much affected the
evolution of $\pp{}$ variables (which have continued to grow). Eventually
however, the exponential growth of $\pmm{}$ terms causes the $V_L$ term
in the equations for $\pp{x}$ and $\pp{y}$, insofar neglible, become
dominant.  Large $V_L$ then forces $\pp{x}$ and $\pp{y}$ to change sign
and grow to opposite direction until $V_L$ again changes the sign.
Additionally, the ensuing oscillatory motion of the $P^+_{x,y}$
components induces the oscillation into other variables as well,
leading to the exponentially large oscillation pattern observed in
$L_{\nu_\tau}$.\\

To find out the extent of chaotic and/or regular behaviour of \sgL,
we have scanned through the parameter space depicted in Fig.\ 2,
which shows the sign of the final asymmetry $L$ with
the initial value $L^{\rm in} = 10^{-10}$.  As can be seen, the
structure of \sgL\ is highly complex.
In the upper left hand corner, extending downwards to large $\theta$,
there is a regular region with no change in the asymmetry. Its
existence is relatively easy to understand: this is the region where
only the very first oscillation is carried out before the sign of
the asymmetry is fixed. Since the direction of the first oscillation
is determined by the sign of the initial asymmetry $L^{\rm in}$ (not
necessarily the initial $\nu_\tau$-asymmetry), the sign of final
neutrino asymmetry in this part of the parameter space should indeed
be regular and fully determined.
\begin{figure}[t]
\centering
\vspace*{-2mm}
\hspace*{-3mm}
\leavevmode\epsfysize=13.0cm \epsfbox{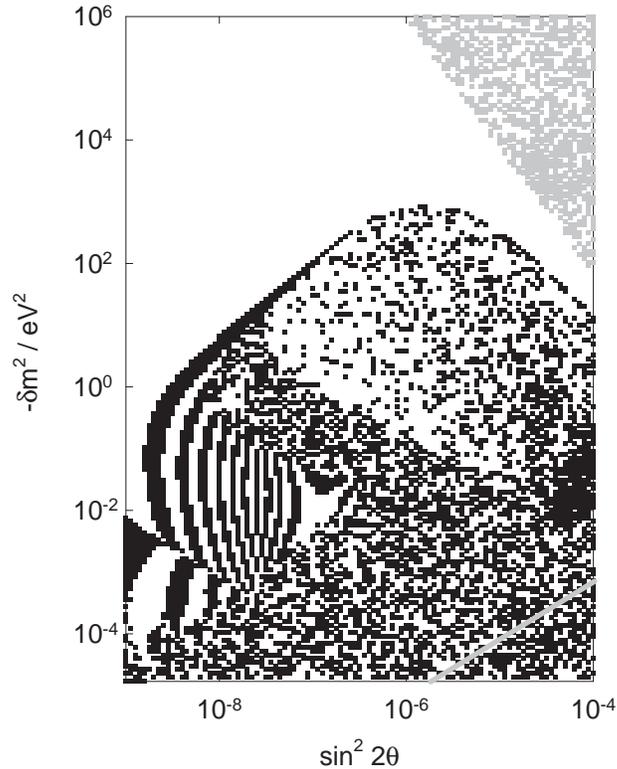}\\[-24mm]
\caption[fig2]{\label{fig2}
         The distribution of the final sign of the neutrino asymmetry in
         the mixing parameter space. Negative \sgL\ is plotted in black,
         positive \sgL\ in white. The initial asymmetry was chosen to be
         $L^{\rm in}= 10^{-10}$.  }
\end{figure}
The bands seen in the left hand side of Fig.\ 2 are formed as the system
goes through two or more oscillations.  In this region the number of
oscillations is slowly increased as $\theta$ grows leading to less
determined \sgL\, but it still can hardly be described as chaotic yet.

In addition to the two more or less regular regions there are regions where
\sgL\ appears to be chaotic. The interval $10^{-2} \lsim |\delta m^2| \lsim
1$ contains a very complicated structure. For $|\delta m^2| \gsim 1$ one may
discern some tendency for positive $L$ to prevail, while the region with
$|\delta m^2| \lsim 10^{-2}$ appears to be pure white noise. In the lower
right-hand corner of Fig. 2 (below the gray line), with
$|\delta m^2|/\sin^2 2\theta \lsim 10$,
oscillations in $L$ will continue past the neutrino freeze-out and will not
settle into any definite value. The boundary of the regular region above which
$L$ is positive is given approximately by
\be
          \left( \log \frac{-\delta m^2}{10^{-5.8} \ev^2} \right)^2
  - 0.44  \left( \log \frac{\sin^2 2\theta}{10^{4.5}} \right)^2 \simeq -1.
\ee

We conjecture that the chaotic behaviour occurs only when the oscillating
period is long. We have also explored in finer detail a restricted region
of parameters in the chaotic regime, without finding any structure. It is
possible however, that the final sign of $L$ in this region is affected by
the accumulated numerical error originating from the extremely high number
of oscillation periods. In this sense proving a true chaoticity is of course
not possible. Nevertheless, if the system is sensitive to numerical error,
it should be expected to be sensitive to parameter fluctuations as well, so
the general pattern of rapid \sgL\ fluctuations is expected to be robust.
Finally, the region in the upper right-hand corner does not correspond to
a large asymmetry, but it is merely the region where $\nu_s$ is fully
equilibrated \cite{ektBig,ektat,ssf} and the absolute value of final
$L$ is very small.

Changing the initial value for $L^{\rm in}$ does not change the picture
qualitatively, although the structures evident in Fig.\ 2 shift slightly to
the left when $L^{\rm in}$ is increased. Moreover, the changes saturate at
$L^{\rm in}\gsim 10^{-9}$.  These changes, or their absence, are very
interesting however: In Fig.\ 3 we have plotted the value of $L$ at the
temperature $T = T_{\rm res} - 2.5$, as a function of $L^{\rm in}$
for three representative choices of parameters. The first set, with
$\sin^22\theta = 10^{-7}$ and $\delta m^2 = -10^4$ eV$^2$ corresponds to
the stable region with positive $L$ in Fig.\ 2.  As expected, $L$ remains
positive independently of the initial value $L^{\rm in}$. In fact the
dependence turned out to be smooth and linear showing that not only is the
sign robust, but also that the numerical solution is well under control.
\begin{figure}[t]
\centering
\vspace*{-7mm}
\hspace*{-3mm}
\leavevmode\epsfysize=13.0cm \epsfbox{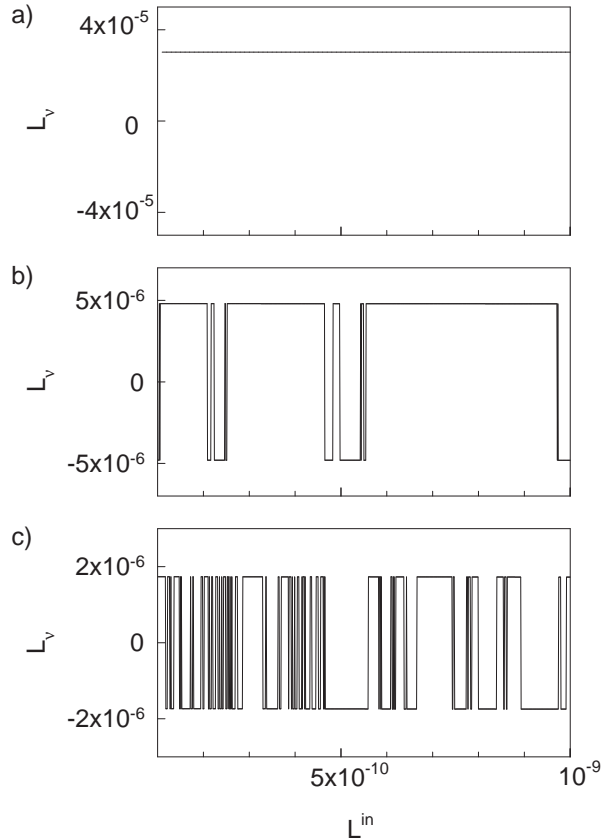}\\[-8mm]
\caption[fig2]{
         The sign of the asymmetry $L_{\nu_\tau}$ at
         $T = T_{\rm res} - 2.5$ MeV
         as a function of the initial asymmetry $L^{\rm in}$ for three sets
         of parameters ($\sin^22\theta$, $\delta m^2/{\rm eV}^2$):
         a) ($10^{-7}$,$-10^4$), b) ($10^{-6}$,$-1$) and
         c) ($10^{-6}$,$-10^{-3}$). }
\end{figure}
The second set, with $\sin^22\theta = 10^{-6}$ and $\delta m^2 = -1$
eV$^2$, lies in the intermediate region where positive $L$ predominates, and
the same dominance is seen as a function of the inital value $L^{\rm in}$.
The last set with $\sin^22\theta = 10^{-6}$ and $\delta m^2 = -10^{-3}$ eV$^2$
corresponds to the chaotic region. It is evident that \sgL\ is very sensitive
to initial conditions, displaying clear randomness as a function of
$L^{\rm in}$. \\

%
%

The final value of \sgL\ has consequences for both atmospheric neutrinos
and primordial nucleosynthesis. It has been proposed that Super-Kamiokande
results for atmospheric neutrinos, which lie in the forbidden zone
\cite{ektBig,ssf,ektat}, might still allow a active-sterile mixing solution
if the asymmetry growth is taken into account \cite{fvA}. Although the
oscillation parameters in the case of atmospheric neutrinos are in the region
where asymmetry growth is not expected, it has been argued that other neutrino
oscillations could induce a large asymmetry in the active-sterile sector which
the oscillations cannot damp.

If the outcome of neutrino oscillations is highly chaotic, the validity of
such a scenario might be suspect. However, we found a large region in the
parameter space where \sgL\ is very robust with respect to small variations
of the mixing parameters. No chaoticity should be expected there with respect
to other small perturbations, such as local perturbations in $L^{\rm in}$,
either.  It is in these stable domains where one would expect that the
mechanism of ref.\ \cite{fvA} can be successful.

In the region where \sgL\ is chaotic in the oscillation parameter space,
it was also found to be sensitive to fluctuations in $L^{\rm in}$; these
are predicted to be generated for example during the QCD phase transition,
or in scenarios of electroweak baryogenesis \cite{qcdpt,baryo,kkss}.
In such case causally disconnected regions would be expected to develop
large asymmetries with a random sign distribution. It has been argued that
then the nucleosynthesis constraint on active-sterile mixing would be
even more stringent, because of additional MSW conversion taking place in
the boundaries of domains with different \sgL\ \cite{shifu2}. However, our
results indicate that the new constraints obtained in \cite{shifu2} may be
overly optimistic, because for a large part of their excluded region we
have found \sgL\ to be stable against small fluctuations; hence in no
domain formation should be expected to occur in the first place.

Determining the sign of $L$ is important also for considering the effect of
the electron (anti)neutrino spectrum distortions on the light element
abundances \cite{eAsymEf,kir}. When the momentum spectrum gets distorted
from its thermal equilibrium value the neutron to proton freezing ratio
will change. Direct $\nu_e \leftrightarrow \nu_s$ oscillations obviously
can induce such distortions, but also scenarios where large asymmetry
is first generated in $\nu_{\mu,\tau} \leftrightarrow \nu_s$ and then
transferred to electron neutrino via $\nu_e \leftrightarrow \nu_{\mu,\tau}$
oscillation, could have considerable effects on the electron neutrino
spectrum. It turns out that positive \sgL\ has the effect of decreasing and
negative \sgL\ of increasing $^4$He abundance \cite{eAsymEf}, so that the
difference is $\Delta Y\simeq 10^{-2}$, with some dependence on the oscillation
parameters. Because the oscillation parameters cannot be measured with an
arbitrary accuracy, it follows that {\em in the region where the \sgL\ is
chaotic, the role of resonant active-sterile neutrino mixing in Big-Bang
Nucleosynthesis can not be reliably estimated}. Rather, in this region,
depicted in Fig.\ 2, there always remains an arbitrariness in the $^4$He
abundance given by  $\Delta Y\simeq 10^{-2}$, which should be considered
as a source of systematic error.

In the region where the sign is stable, more concrete conclusions can be
drawn. However, in this region $L$ is positive and only a rather small
negative shift in the helium abundance  $-0.005 \lsim \Delta Y < 0$ was
found \cite{sfa} for these parameters. Interestingly enough, such a
shift could ameliorate the apparent conflict of the nucleosynthesis
theory viz-a-viz observations \cite{RecNS}.

Our results in this paper are based on an averaged momentum description
of the neutrino ensemble. Some effects, like the diffusion of the asymmetry
between different momentum states, would seem to indicate the need for using
full momentum-dependent kinetic equations. This is rather hard, since one has
to deal with exponential growth in every momentum state and the width of the
resonance is, for most of the parameter space, so small that one needs a very
large number of bins to complete the task.  Our preliminary results with
momentum dependent kinetic equations support the results presented here. \\

%
%
\noindent This work has been supported by the Academy of Finland  under
the contract 101-35224.
%
%
\nc{\advp}[3]{{\it  Adv.\ in\ Phys.\ }{{\bf #1} {(#2)} {#3}}}
\nc{\annp}[3]{{\it  Ann.\ Phys.\ (N.Y.)\ }{{\bf #1} {(#2)} {#3}}}
\nc{\apl}[3] {{\it  Appl. Phys. Lett. }{{\bf #1} {(#2)} {#3}}}
\nc{\apj}[3] {{\it  Ap.\ J.\ }{{\bf #1} {(#2)} {#3}}}
\nc{\apjl}[3]{{\it  Ap.\ J.\ Lett.\ }{{\bf #1} {(#2)} {#3}}}
\nc{\app}[3] {{\it  Astropart.\ Phys.\ }{{\bf #1} {(#2)} {#3}}}
\nc{\cmp}[3] {{\it  Comm.\ Math.\ Phys.\ }{{ \bf #1} {(#2)} {#3}}}
\nc{\cqg}[3] {{\it  Class.\ Quant.\ Grav.\ }{{\bf #1} {(#2)} {#3}}}
\nc{\epl}[3] {{\it  Europhys.\ Lett.\ }{{\bf #1} {(#2)} {#3}}}
\nc{\ijmp}[3]{{\it  Int.\ J.\ Mod.\ Phys.\ }{{\bf #1} {(#2)} {#3}}}
\nc{\ijtp}[3]{{\it  Int.\ J.\ Theor.\ Phys.\ }{{\bf #1} {(#2)} {#3}}}
\nc{\jmp}[3] {{\it  J.\ Math.\ Phys.\ }{{ \bf #1} {(#2)} {#3}}}
\nc{\jpa}[3] {{\it  J.\ Phys.\ A\ }{{\bf #1} {(#2)} {#3}}}
\nc{\jpc}[3] {{\it  J.\ Phys.\ C\ }{{\bf #1} {(#2)} {#3}}}
\nc{\jap}[3] {{\it  J.\ Appl.\ Phys.\ }{{\bf #1} {(#2)} {#3}}}
\nc{\jpsj}[3]{{\it  J.\ Phys.\ Soc.\ Japan\ }{{\bf #1} {(#2)} {#3}}}
\nc{\lmp}[3] {{\it  Lett.\ Math.\ Phys.\ }{{\bf #1} {(#2)} {#3}}}
\nc{\mpl}[3] {{\it  Mod.\ Phys.\ Lett.\ }{{\bf #1} {(#2)} {#3}}}
\nc{\ncim}[3]{{\it  Nuov.\ Cim.\ }{{\bf #1} {(#2)} {#3}}}
\nc{\np}[3]  {{\it  Nucl.\ Phys.\ }{{\bf #1} {(#2)} {#3}}}
\nc{\pr}[3]  {{\it  Phys.\ Rev.\ }{{\bf #1} {(#2)} {#3}}}
\nc{\pra}[3] {{\it  Phys.\ Rev.\ A\ }{{\bf #1} {(#2)} {#3}}}
\nc{\prb}[3] {{\it  Phys.\ Rev.\ B\ }{{{\bf #1} {(#2)} {#3}}}}
\nc{\prc}[3] {{\it  Phys.\ Rev.\ C\ }{{\bf #1} {(#2)} {#3}}}
\nc{\prd}[3] {{\it  Phys.\ Rev.\ D\ }{{\bf #1} {(#2)} {#3}}}
\nc{\prl}[3] {{\it  Phys.\ Rev.\ Lett.\ }{{\bf #1} {(#2)} {#3}}}
\nc{\pl}[3]  {{\it  Phys.\ Lett.\ }{{\bf #1} {(#2)} {#3}}}
\nc{\prep}[3]{{\it  Phys.\ Rep.\ }{{\bf #1} {(#2)} {#3}}}
\nc{\prsl}[3]{{\it  Proc.\ R.\ Soc.\ London\ }{{\bf #1} {(#2)} {#3}}}
\nc{\ptp}[3] {{\it  Prog.\ Theor.\ Phys.\ }{{\bf #1} {(#2)} {#3}}}
\nc{\ptps}[3]{{\it  Prog\ Theor.\ Phys.\ suppl.\ }{{\bf #1} {(#2)} {#3}}}
\nc{\physa}[3]{{\it Physica\ A\ }{{\bf #1} {(#2)} {#3}}}
\nc{\physb}[3]{{\it Physica\ B\ }{{\bf #1} {(#2)} {#3}}}
\nc{\phys}[3]{{\it  Physica\ }{{\bf #1} {(#2)} {#3}}}
\nc{\rmp}[3] {{\it  Rev.\ Mod.\ Phys.\ }{{\bf #1} {(#2)} {#3}}}
\nc{\rpp}[3] {{\it  Rep.\ Prog.\ Phys.\ }{{\bf #1} {(#2)} {#3}}}
\nc{\sjnp}[3]{{\it  Sov.\ J.\ Nucl.\ Phys.\ }{{\bf #1} {(#2)} {#3}}}
\nc{\sjp}[3] {{\it  Sov.\ J.\ Phys.\ }{{\bf #1} {(#2)} {#3}}}
\nc{\spjetp}[3]{{\it Sov.\ Phys.\ JETP\ }{{\bf #1} {(#2)} {#3}}}
\nc{\yf}[3]  {{\it  Yad.\ Fiz.\ }{{\bf #1} {(#2)} {#3}}}
\nc{\zetp}[3]{{\it  Zh.\ Eksp.\ Teor.\ Fiz.\ }{{\bf #1} {(#2)} {#3}}}
\nc{\zp}[3]  {{\it  Z.\ Phys.\ }{{\bf #1} {(#2)} {#3}}}
\nc{\ibid}[3]{{\sl  ibid.\ }{{\bf #1} {#2} {#3}}}
%
%

\end{document}